\documentclass[letterpaper, 10 pt, conference]{ieeeconf}  

\IEEEoverridecommandlockouts                              

\usepackage{xcolor,soul}
\usepackage{amsmath,graphicx, siunitx, dblfloatfix, booktabs}
\usepackage{balance}
\usepackage{placeins}
\usepackage[flushleft]{threeparttable}
\usepackage{adjustbox}

\title{\LARGE \bf
Segmentation-free Heart Pathology Detection Using Deep Learning}

\author{Erika Bondareva$^{1}$, Jing Han$^{1}$,  William Bradlow$^{2}$, Cecilia Mascolo$^{1}$

\thanks{$^{1}$The Department of Computer Science and Technology, University of Cambridge
        {\tt\small eb729@cam.ac.uk}}%
        
\thanks{$^{2}$Department of Cardiology, University Hospitals Birmingham NHS Foundation Trust, 9 Birmingham and
Institute of Cardiovascular Sciences, University of Birmingham.}%

\thanks{*This work was supported by the UK Engineering and Physical Sciences Research Council (EPSRC) grant EP/L015889/1 for the Centre for Doctoral Training in Sensor Technologies and Applications, and by ERC Project 833296 (EAR).}
}

\begin{document}

\maketitle
\thispagestyle{empty}
\pagestyle{empty}

\begin{abstract}
Cardiovascular (CV) diseases are the leading cause of death in the world, and auscultation is typically an essential part of a cardiovascular examination. 
The ability to diagnose a patient based on their heart sounds is a rather difficult skill to master. Thus, many approaches for automated heart auscultation have been explored. However, most of the previously proposed methods involve a segmentation step, the performance of which drops significantly for high pulse rates or noisy signals. 
In this work, we propose a novel segmentation-free heart sound classification method. Specifically, we apply discrete wavelet transform to denoise the signal, followed by feature extraction and feature reduction. 
Then, Support Vector Machines and Deep Neural Networks are utilised for classification. 
On the PASCAL heart sound dataset our approach showed superior performance compared to others, achieving 81\% and 96\% precision on normal and murmur classes, respectively. In addition, for the first time, the data were further explored under a user-independent setting, where the proposed method achieved 92\% and 86\% precision on normal and murmur, demonstrating the potential of enabling automatic murmur detection for practical use.

\end{abstract}

\section{INTRODUCTION}

\label{sec:intro}

Cardiovascular (CV) diseases are the number one cause of mortality internationally, accounting for 17 million deaths per year according to the World Health Organisation~\cite{WHOcvd}.
Auscultation, the process of listening to the heart using a stethoscope, is a powerful screening tool due to its affordability, non-invasiveness, safety, and ease of administration; but it is a very difficult skill to master. Therefore, researchers are starting to explore the potential of digital technologies for both sound gathering and automatic data analysis. 

To fully appreciate the capabilities of CV auscultation, it is important to understand the origin of the sounds. One full cardiac cycle consists of two sounds, commonly referred to as S1 and S2, which are caused by valve closure. Occasionally,
there are additional sounds present, which are most often indicative of a pathology.


Conventional heart sound classification methods typically involve sound filtering, segmentation, feature extraction, and machine learning-based inference. However, the performance on a wider sample population is rarely a concern. It has been reported that many denoising approaches in the literature do not work on real-life data due to the algorithms developed on artificially noisy data~\cite{Gradolewski2014}. Additionally, there are questions and issues remaining which are discussed below and investigated particularly in this study.

\textit{Limitations of segmentation:}
One of the most common segmentation categories are envelope-based methods, including normalised Shannon average energy~\cite{Gomes2012, Balili2015, Chakir2018} and Hilbert transform and its variations~\cite{Huang1998, Cherif2013, Sharma2015}. These methods, however, are based on the premise that systole is shorter than diastole, which is only true for a narrow band of heart rates~\cite{Deng2012}, and they become less effective for noisy signals~\cite{Dwivedi2019, Asmare2020}. As a consequence, segmentation's reliance on the signal being clean within a narrow band of heart rates drastically limits its application, with the error introduced by incorrect segmentation propagating into heart sound pathology classification, and introducing considerable computational complexity into the algorithms~\cite{Langley2017}. 
To avoid such complexity and the underlying error propagation issue, in this work we propose a segmentation-free framework for heart sound classification. 
To the best of our knowledge, there are very few segmentation-free 
methods~\cite{deng2016towards}, and our method shows better performance.

\textit{Lack of performance comparison:}
To tackle the heart sound classification problem, a variety of models have been investigated. For instance, Support Vector Machines (SVMs) are arguably the most popular classifier owing to their mature theoretical foundation, with a plethora of research attempts to explore various feature inputs for heart sound classification. Some utilised SVMs with deep features as inputs, which yielded a sensitivity of 84.8\%~\cite{Tschannen2016}, and others applied sparse coding for feature extraction, achieving 86.5\% accuracy~\cite{Whitaker2016}, both on PhysioNet heart sound dataset. 
More recently, Deep Neural Networks (DNNs) have started to receive greater attention for this task~\cite{Olmez2003, Nilanon2016, noman2019}. While DNNs deal exceptionally well with multi-class problems, unlike SVMs, they require a large amount of training data. However, owing to the lack of fair performance comparison, no clear observations can be drawn as to the superiority of any of them. In our study, we compare the inferential capabilities of SVMs and DNNs on the same data set, and the results indicate that DNNs appear superior to SVMs for heart sound classification in most cases.



\textit{Poor reproducibility issue:}
Concerns about poor reproducibility of digital technologies have been raised in a variety of research areas, including biomedical engineering. In particular, the original PASCAL heart sound challenge~\cite{pascal} divided the data into train and test sets, but user-independent constraint was not maintained: two different sound samples from the train and test sets may had been collected from the same patient. In the present study, we address this issue and re-evaluate the performance of our framework with user-independent constraints. This simulates the utmost authentic assessment in clinical practice and may increase the generalisability of our findings. Our results demonstrate that the proposed approach is able to detect heart murmurs from an unseen patient with 86\% precision via a DNN model.




Motivated by previous encouraging studies and trying to tackle the aforementioned understudied issues for heart sound classification, in this work we propose a novel framework for heart sound classification. In particular, we focus on detecting heart murmurs among normal heart sounds and sounds with extrasystole. We develop a segmentation-free approach for the task at hand. Also, we carry out extensive experiments on the PASCAL heart sound dataset, aiming at fairly comparing the effectiveness and robustness of SVM and DNN models. More importantly, we evaluate the performance under a user-independent setting to verify our model's generalisability to heart sound data from unseen users. Our results show that the proposed approach outperforms other state-of-the-art methods.

\section{METHODOLOGY}

\subsection{Data and its limitations}

The PASCAL Classifying Heart Sounds Challenge dataset (2011)~\cite{pascal} is the first large publicly available heart sound dataset. We used the PASCAL heart sound dataset B for this study, which was collected using a stethoscope. Dataset B contains three classes: normal, murmur, and extrasys (for extrasystole). Each class contains audio recordings of heart sounds, further referred to as samples. The labelled samples are split into two subgroups -- noisy and ``clean'' (comparatively) samples. There are 149 noisy samples: 120 normal and 29 murmur, and a total of 461 clean samples: 200 normal samples, 66 with a murmur, and 46 with extrasys. Also, there are 195 unlabelled sound samples for testing. However, the user identification information can still be fetched from the unlabelled data, and thus can be re-labelled and exploited to enlarge the size of the data for training.


Worth mentioning that the patients whose samples are labelled as extrasys also have other samples in the dataset that do not have the extrasystole present and therefore are normal. Extrasys is essentially a normal heart sound, except at some point during the recording of any length there is an additional heart sound present. 
This knowledge was leveraged for data manipulation, described in section~\ref{dataset-manipulation}.

\subsection{Data preprocessing}

\subsubsection{Denoising}

Heart sound denoising was performed using a Finite Impulse Response (FIR) high-pass filter followed by a Discrete Wavelet Transform (DWT), utilising MATLAB’s Wavelet Toolbox.
The high-pass filtering removed all frequencies below \SI{60}{\hertz}. 
Various parameters were tested for wavelet denoising, and the parameters that resulted in the removal of the most noise while keeping the rest of the signal intact were selected and applied to the whole dataset. Specifically, Daubechies 4 was selected as the mother wavelet, 6$^\text{th}$ decomposition level was used, heursure threshold selection method,  
hard thresholding, and threshold rescaling using level-dependent estimation of level noise.
Finally, before saving the denoised audio samples for feature extraction and classification, the signal was normalised in the range between -0.5 and 0.5, and centred so that the mean is equal to zero.


\subsubsection{Feature extraction}

The INTERSPEECH ComParE 2018 feature set (IS-18)~\cite{schuller2018interspeech} 
was extracted from the signal, resulting in a vector with 6373 features.
This feature set has been applied successfully for many acoustic tasks.
In order to enable extrasys detection, additional features were computed. These features were based on Shannon energy ($E_{\text{Shannon}}$) calculated for the signal using the equation:
\begin{equation}
    E_{\text{Shannon}} = -(x ^{2}) * log_{10} x ^{2},
\end{equation}
where $x$ denotes the signal. Then, a moving average was calculated with 100 samples window length. The resulting Shannon energy envelope was normalised, and peaks were extracted using a height threshold of 0.08 and distance of 400 samples, and the peak-to-peak distance was calculated. Finally, a number of features was extracted based on the peak-to-peak distance as well as peak timing: two largest and two smallest values, as well as a mean and a standard deviation. In addition, the total number of peaks was counted. The resulting 13 features were concatenated into a vector, and appended to the IS-18 feature vector. 

While this approach still includes peak detection, it does not rely on the correct classification of S1 and S2 heart sounds, as the methods presented for extrasys detection tend to. Therefore, this approach is likely to be more robust to noise inherently captured with heart sounds.

\subsubsection{Feature reduction}
The resulting feature vector had 6386 components. The extracted features were standardised by removing the mean and scaling to unit variance, and the standardised features were then reduced by utilising principal component analysis (PCA). 460 components were retained, with variance ratio 99.99\%.

\subsection{Classification}

Python Tensorflow and Keras libraries were used for designing the classifiers and evaluating their performance. We decided to apply an SVM (with an RBF kernel) and DNN for classification, given the success reported in previous works.
For the DNN, the design that yielded the best result was a 6-layer neural network, with 512 nodes in the first two layers, 256 in the middle two, and 128 nodes in the last two layers. We employed ReLU activation,
the dropout 0.2 in the first layer and 0.5 on all the following layers,
and Adam optimiser.
The model was trained for a varying number of epochs, depending on the experiment, and the numbers are specified and explained in section~\ref{experiment-design}.

\subsection{Evaluation}

The PASCAL challenge suggested a number of specific metrics for evaluation of the models:
\begin{itemize} 
	\item \textbf{Precision} per individual class --- a measure of how many samples belonging to a specific class are labelled correctly, calculated as $ \text{TP} / (\text{TP} + \text{FP}) $. When reporting results, PN, PM, and PE stand for precision of normal, murmur, and extrasys respectively.
	\item \textbf{Sensitivity} ($\text{Sens} =  \text{TP} / (\text{TP} + \text{FN})$) and \textbf{specificity} ($\text{Spec} = \text{TN} / (\text{TN} + \text{FP})$).
	\item \textbf{Youden's index} ($\gamma$) 
	--- a metric demonstrating the test's ability to avoid failure, which can be calculated using $
	\gamma = \text{Sens} + \text{Spec} - 1$.
	\item \textbf{Discriminant power} ($\text{D}$) also summarises sensitivity and specificity:  
	\begin{equation*}
	\text{D} = \frac{\sqrt{3}}{\pi} \bigg( \log \bigg( \frac{\text{Sens}}{1 - \text{Sens}}  \bigg) + \log \bigg(  \frac{\text{Spec}}{1 - \text{Spec}} \bigg)  \bigg)
	\end{equation*}
	\item \textbf{Total precision} --- the sum of precision values obtained per individual class. Maximum total precision for three-class problem is $3.0$, and $2.0$ for binary classification.
\end{itemize}

\subsection{Dataset manipulation}
\label{dataset-manipulation}
The dataset was manipulated for some of the experiments conducted as a part of this work, due to the concern that the split between train and test set suggested by the PASCAL challenge does not allow to perform cross-validation or user-independent testing. 
Keeping in mind the superior clinical importance of murmur detection, all extrasys files were relabelled into normal (or, rather, sounds with no murmur present), and the patient identification number was used to label all the unlabelled samples from the test set. 

\subsection{Experiment design}
\label{experiment-design}

Three experiments were designed for this study:
\begin{itemize}
\item \textbf{Exp 1 (Three-class Classification)} --- trained and evaluated the model using the PASCAL dataset the way the challenge was intended: training on three classes (normal, murmur, and extrasys), and evaluating using the included locked spreadsheet that automatically calculates the suggested evaluation metrics. The deep learning model was trained for 500 epochs. Worth noting, that the train/test split is not user-independent, but the experiment was performed for method comparison with previous results reported on the PASCAL dataset.
\item\textbf{Exp 2 (Binary Classification)} --- relabelled all the extrasys samples to normal.
As a result, instead of three-class classification, a \textit{binary classification} (normal and murmur) was performed. 
Then, we also labelled all the unlabelled samples from the test set using the patient ID under the assumption that each patient has either normal or murmur heart sounds. Given that the classification problem became binary, only 50 epochs were required to train the DNN model. The main goal of this experiment was to observe the effect the increased amount of data has on the classifier performance.
\item \textbf{Exp 3 (User-Independent)} --- used the same data and labels as in Exp 2, but with \textit{user-independent} train/test split. The DNN model benefited from a slight increase in the number of epochs to 100. This experiment aimed to investigate the classifier performance in a more realistic setting, on previously unseen users. 
\end{itemize}
\vspace{-0.15cm}
For all of the experiments, both SVM and DNN classifiers were used for performance comparison. 

Furthermore, for Exp1, we compared our approach with other state-of-the-art methods reported on the PASCAL dataset.
In particular, the two winners of the challenge proposed employing segmentation-based features, with the first group utilising J48 and multi-layer perceptron (MLP) algorithms~\cite{Gomes2012}, while the second group proposed an algorithmic approach (no machine learning)~\cite{Deng2012}. In addition, discrete and continuous wavelet transform coefficients were explored as inputs to a random forest classifier in~\cite{Balili2015}. Segmentation-based features and a discriminant analysis classifier were used in~\cite{Chakir2018}. Last, our approach was also compared with another segmentation-free method, which utilised autocorrelation features and diffusion maps~\cite{deng2016towards}.

\section{RESULTS AND DISCUSSION}
The proposed methodology maintains high performance on the original challenge, while offering a significant improvement on the detection of murmur. It does not employ segmentation, which is highly desirable, considering segmentation's susceptibility to noise and elevated heart rates. In addition, we demonstrate improved performance across all metrics by increasing the amount of data in the training set, managing to preserve the high performance for the user-independent split.


The results of Exp 1 and the comparison to other state-of-the-art results on the PASCAL dataset are presented in Table 1. DNN demonstrates better performance than the SVM, and the segmentation-free methodology presented in this paper yields superior results for the following metrics: the precision of murmur, the specificity of heart problem, Youden index of heart problem, and total precision. The metrics outperformed by other approaches are precision of normal, where our achieved precision is only marginally worse (1\% reduction), the precision of extrasys, and discriminant power.

\begin{table}[t!]
	\vspace{0.2cm}
	\caption{Results of Exp 1.}
	\label{tab:exp1_results}
	\centering
\begin{threeparttable}
	\begin{tabular}{ l | c c c c c | c c}
		\toprule
		&
		\multicolumn{5}{c|}{\textbf{Previous works}} &
		\multicolumn{2}{c}{\textbf{Our method}}
		\\
		&
        \textbf{\cite{Gomes2012}} &
		\textbf{\cite{Deng2012}} &
		\textbf{\cite{Balili2015}} &
		\textbf{\cite{Chakir2018}} &
		\textbf{\cite{deng2016towards}} &
		\textbf{SVM} &
		\textbf{DNN} \\ \midrule 
		PN &
		0.70 &
		0.77 &
		0.71 &
        \textbf{0.82} &
        0.77 &
        \textbf{0.82} &
		0.81 \\ 
		PM &
		0.30 &
		0.37 &
		0.33 &
		0.59 &
		0.76 &
		0.70 &
         \textbf{0.96} \\ 
		PE &
		0.67 &
		0.17 &
        \textbf{1.00} &
		0.18 &
		0.50 &
		0.20 &
		0.50 \\ 
		Sens &
		0.19 &
		0.51 &
		0.14 &
		0.49 &
		0.34 &
		\textbf{0.54} &
		0.47 \\ 
		Spec &
		0.84 &
		0.59 &
		0.90 &
		0.66 &
		0.95 &
		0.77 &
		\textbf{0.99} \\ 
		{$\gamma$} &
		0.02 &
		0.01 &
		0.04 &
		0.15 &
		0.29 &
		0.31 &
		\textbf{0.46} \\ 
		D &
		0.04 &
		0.09 &
		0.09 &
		0.15 &
		\textbf{1.24} &
		0.33 &
		0.98 \\ 
		TPr &
		1.67 &
		1.31 &
		2.04 &
		1.58 &
		2.03 &
		1.72 &
		\textbf{2.28}
		\\ \bottomrule
	\end{tabular}
\end{threeparttable}
\vspace{-0.2cm}
\end{table}
Exp 2 and 3 involved binary classification and increasing the amount of training data, which, as expected, resulted in increased performance (see Table 2). Worth noting that the results of the user-independent train/test split (Exp 3) yielded slightly poorer results, which is expected. 

Comparing two classifiers, it could be argued that the deep learning model yields better performance in comparison to the SVM. While for the user-independent split SVM demonstrates higher total precision and lower standard deviation across five folds, DNN yields significantly higher sensitivity of the heart problem. 

\begin{table}[t!]
\vspace{0.15cm}
	\caption{Results of Exp 2 and 3. For each evaluation metric its mean and st dev are reported across five-fold cross-validation.}
	\label{tab:exp234_results}
	\centering
\begin{threeparttable}
\begin{tabular}{ l | c c | c c }
	\toprule &
    \multicolumn{2}{c|}{\textbf{Exp 2}} &
    \multicolumn{2}{c}{\textbf{Exp 3}}  \\ 
    &
    \textbf{SVM} & \textbf{DNN} & \textbf{SVM} & \textbf{DNN}  \\ \midrule 
	{PN}        & 0.88$\pm$0.01   & 0.92$\pm$0.03   & 0.88$\pm$0.01 & 0.92$\pm$0.02  \\ 
	{PM}        & 0.99$\pm$0.03   & 0.97$\pm$0.04   & 0.98$\pm$0.04 & 0.86$\pm$0.11  \\ 
	{Sens}      & 0.50$\pm$0.07   & 0.68$\pm$0.12   & 0.48$\pm$0.06 & 0.68$\pm$0.07  \\ 
	{Spec}      & 1.00$\pm$0.00   & 0.99$\pm$0.01   & 1.00$\pm$0.00 & 0.97$\pm$0.03  \\ 
	{$\gamma$}  & 0.50$\pm$0.07   & 0.67$\pm$0.12   & 0.48$\pm$0.07 & 0.65$\pm$0.09  \\ 
	{D}         & N/A               & N/A               & N/A           & N/A  \\ 
	{TPr}        & 1.87$\pm$0.04   & 1.89$\pm$0.04   & 1.86$\pm$0.05 & 1.78$\pm$0.13  \\
 \bottomrule
 
\end{tabular}
\begin{tablenotes}
      \small
      \item It was impossible to calculate the discriminant power $D$ across the five folds since in some folds specificity of murmur was $1$, resulting in subsequent division by zero.
    \end{tablenotes}
\vspace{-2em}
\end{threeparttable}
\end{table}

The best performance in terms of the sensitivity of heart problem on user-independent training and testing set splits (Exp 3) was achieved by a DNN classifier, achieving 92\% precision of normal and 86\% precision of murmur, with 68\% sensitivity of murmur.

Total precision slightly dropped for user-independent testing. This suggests that in previous works on the PASCAL dataset, which used the dataset with the data from the same patient appearing in both training and testing sets, according to the original design, inference models might have suffered from overfitting and might yield poorer performance on new samples from previously unseen patients, as a consequence of lacking generalisability and reproducibility. 

While the proposed method achieves superior sensitivity of heart problem in comparison to all the approaches reported to date, it still could be deemed insufficient, given the importance of detection of heart murmurs. It is worth keeping in mind that in the PASCAL dataset every sample coming from the same patient with a murmur receives the same label, despite the murmur amplitude variation across auscultatory locations. This becomes especially problematic in patients with mild murmurs, where in distant locations murmur amplitude does not exceed the noise amplitude, masking the pathology, and thus needs further investigation.

\section{CONCLUSION}
This paper presents a novel segmentation-free method for murmur and extrasystole detection in heart sounds. 
To the best of our knowledge, the proposed framework achieved the best performance on the PASCAL dataset. In addition, upon increasing the training set by manipulating the dataset and implementing user-independent testing to meet the practical needs, we reliably boost the performance by all the evaluation metrics. 


Automated auscultation could become an exceptionally valuable tool for diagnostics and disease progression tracking, but the lack of publicly available granular heart sound datasets limits the research efforts in this field. The directions for future work would include expanding the models to more detailed datasets, as well as considering all the data from a single patient combined to reduce the effect auscultatory location has on the model prediction.




\balance

\bibliographystyle{IEEE}
\bibliography{root}

\end{document}